# GPU-Based Interactive Visualization of Billion Point Cosmological Simulations

Tamas Szalay[1], Volker Springel[2], Gerard Lemson[3,4]

1. California Institute of Technology, California Ave, 91126, Pasadena, CA
2. Max-Planck-Institute for Astrophysics, Karl-Schwarzschild-Str. 1, D-85741 Garching, Germany
3. Astronomisches Rechen-Institut, Zentrum für Astronomie der Universität Heidelberg, Moenchhofstr. 12-14, 69120, Heidelberg, Germany
4. Max-Planck-Institut für extraterrestrische Physik, Giessenbach Str., 85748, Garching, Germany

*Despite the recent advances in graphics hardware capabilities, a brute force approach is incapable of interactively displaying terabytes of data. We have implemented a system that uses hierarchical level-of-detailing for the results of cosmological simulations, in order to display visually accurate results without loading in the full dataset (containing over 10 billion points). The guiding principle of the program is that the user should not be able to distinguish what they are seeing from a full rendering of the original data. Furthermore, by using a tree-based system for levels of detail, the size of the underlying data is limited only by the capacity of the IO system containing it.*

Over the past few years, the quantity of scientific data produced has increased dramatically. This increase has been matched by a corresponding growth in storage capacity and processing power, and many new designs are emerging to deal with the efficient utilization of CPUs and disks. One of the most recent advances is the advent of general-purpose GPUs, which allow orders of magnitude better performance for certain algorithms.

One problem that these systems do not solve, however, is interactive rendering. In fact, even non-interactive rendering is prohibitively expensive, when a three-minute movie can take over a week to render on a computing cluster. If the data to be rendered is multiple terabytes, there is no way to avoid scanning through the entire dataset to produce a perfect rendering.

This is acceptable if one knows exactly what to look for, but this is hardly ever the case. To examine the collision of two small galaxies in a simulation containing millions of them, one would have to painstakingly pore over the data looking for such an event before being able to set up the correct rendering, a tedious process that could take weeks.

The data under consideration comes from the Aquarius simulations, n-body gravitational simulations of dark matter particles. In the version dealt with by the program, there are 150 million particles in each timestep and 128 timesteps in all, giving over 20 billion total particles. The full size of this simulation takes up 1.4 terabytes on disk[1].

So, there is a clear need for some sort of interactive rendering, even for the largest simulations and datasets. Analysis programs like MATLAB and Mathematica have done a good job of keeping up to date and utilizing graphics hardware acceleration, but the bottleneck still lies in the data transfer and volume. The only way to render terabytes of data without scanning the entire set is to draw a visually equivalent version represented by much less data.

The critical steps are figuring out some way to load in only the necessary data, and generating visually equivalent versions of the original data at several scales. This has been done many times for simpler data formats; one need only look at satellite images in Google Maps for a good example.

For cosmological simulations, the visualization procedure generally consists of taking a large number of points (currently up to tens of billions) and rendering each point as a pixel or a group of pixels on the screen. Given that a screen has only around 1 million pixels, it becomes readily apparent that drawing 10 billion points will have significant redundancy. Furthermore, if only the nearest two galaxies are the target of the visualization, maybe only a few million points need to be resolved in significant detail and the rest can be ignored.

So, by precomputing such levels of detail at different scales, it is possible to actually load in a few gigabytes of data and produce output visually equivalent to the original. Of course, as the viewpoint changes, the loaded data will need to change as well, but only at a pace that can easily be met by today's IO speeds. In fact, the rendering hardware could be connected directly to a database or RAID array to get throughputs in GB/s.

More importantly, though, multiple users could connect to the same central storage server, and each of them could run the visualization software on an average desktop computer. A transfer speed of 100 MB/s is enough to dynamically load in the required data, and the desktop computer would only need a $200 graphics card to be able to render it. The data only needs to be provided as fast as the program requests it.

There are three requirements to be able to create a system as described above:

a) a process to simplify the original data to produce a visually equivalent output at a reduced scale
b) spatial indexing on the levels of detail, to be able to load in data as needed
c) a rendering application that can determine what data should be loaded and drawn in detail

The fundamental data structure we used to index the cosmological data is a spatial octree. The constraint is that all the data contained within a spatial node or leaf should be represented by at most n points (in our case, n=16000). If there are more points in the original data, they are simplified using a density-weighted random subselection algorithm until the same data is represented by only n points. Thus, the children of any node contain (spatially) the same data as the parent but in 8x more detail.

Using this structure, it is a simple matter to dynamically determine what data to load: locations close to the viewer will be loaded from deep in the tree, at very high detail, while those far away will be shallow and will display a greatly simplified version of the data. This is the key step – with that, 10 million points can be drawn on screen to represent 10 billion points' worth of data.

A spatial octree organized into such "blocks" of data confers other advantages as well. Most scientific data is stored as double precision, which is of course 8 bytes per coordinate of position. We have quantized the positions of each point to 2 bytes per coordinate, spanning the containing block, but since the blocks get smaller the deeper they are in the tree, the accuracy also increases. As a result, there is no loss of information, but the required amount of data can be read four times as fast.

All of the other properties of the points are encoded in the same way, and this is where the power of the GPU comes into play. The front-end rendering program uses the CPU only for deciding which blocks to load and render; all of the other work is done on the graphics card. Each point also carries time information, allowing the graphics card to interpolate position, size, and color for every frame. Each point is rendered as a smeared blob, and a radial cubic function is computed for every single pixel of every point without any performance impact.

In fact, graphics cards are so amazingly fast that they defy all expectations regarding computational power. As an example, the current program has the ability to highlight specific points by their IDs, a unique number contained in each point. This is done by storing a list of "selected" point IDs in the graphics card, a list up to one million IDs in length, and testing whether each point is a member of that list as it is rendered. On a CPU, this would be a computational disaster; on a GPU, however, this incurs zero performance penalty.

Of course, while all of these ideas have only been applied to the Aquarius data, they could just as easily be applied to other spatial data such as laser scans[2].

Even though modern scientific visualization and analysis tools can render a limited amount of data quite fast, the brute-force approach they use does not scale. By preprocessing the data into hierarchical levels of detail, a visually-accurate substitute can be rendered with far less data. The system we have implemented allows one to:

- View the data at different scales and in different locations, instantly
- Examine the data at different times in the simulation, or start a continuous playback, as the data is streamed from disk
- Select some points of interest and follow them from the beginning of the simulation to the end

The entire front-end is on an average desktop computer with a GeForce 8800 graphics card, and it can do all of the above at 10 frames per second, even when the underlying data is a terabyte in size.

In short, a system such as this allows researchers to interact with large datasets in a way never before possible, hiding the usual clumsiness of manipulating so much data.

Gerard Lemson works for the German Astrophysical Virtual Observatory (GAVO), which is supported by a grant from the German Federal Ministry of Education and Research (BMBF) under contract 05 AC6VHA.

The following screenshots are taken directly from the rendering program.

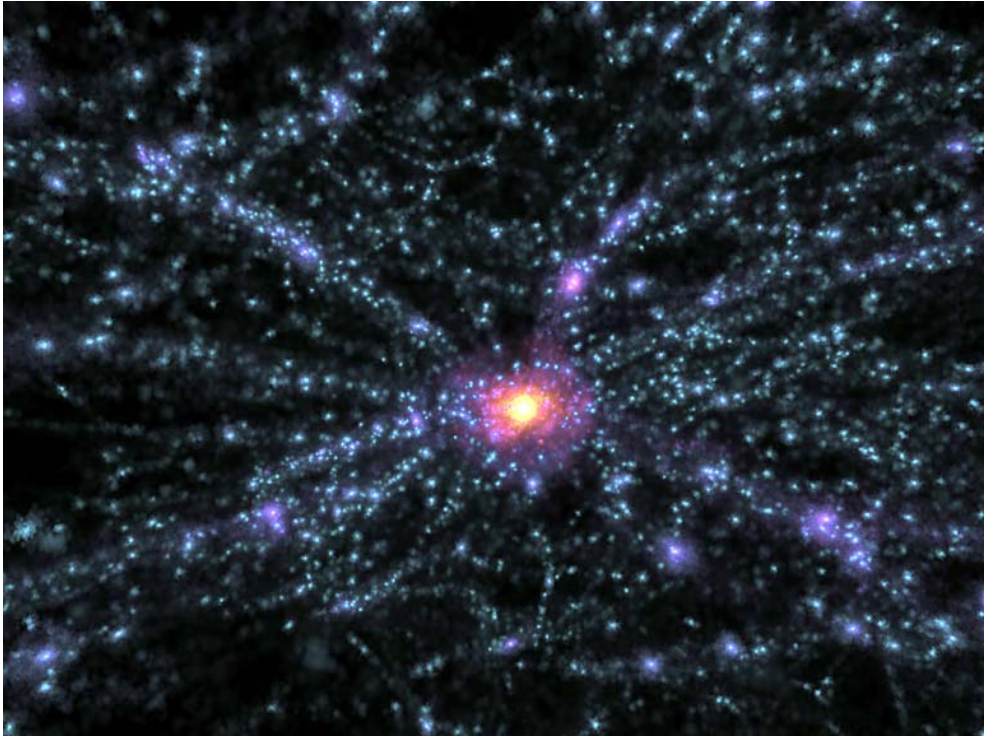

1. An example of normal rendering.

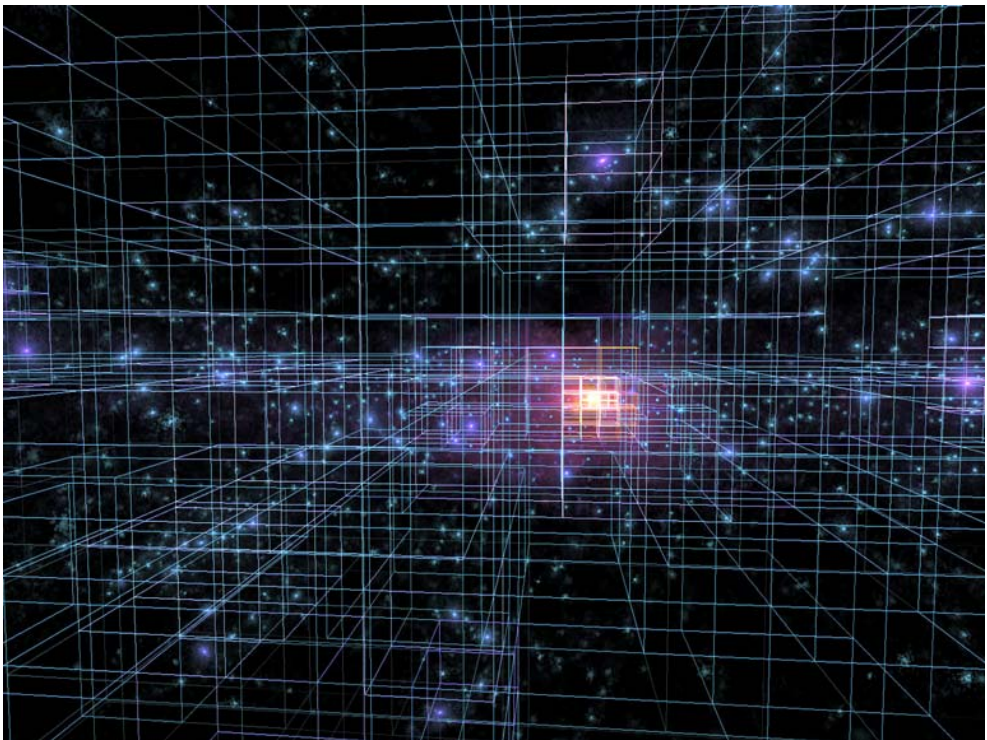

2. The octree structure, each cube is a "block"

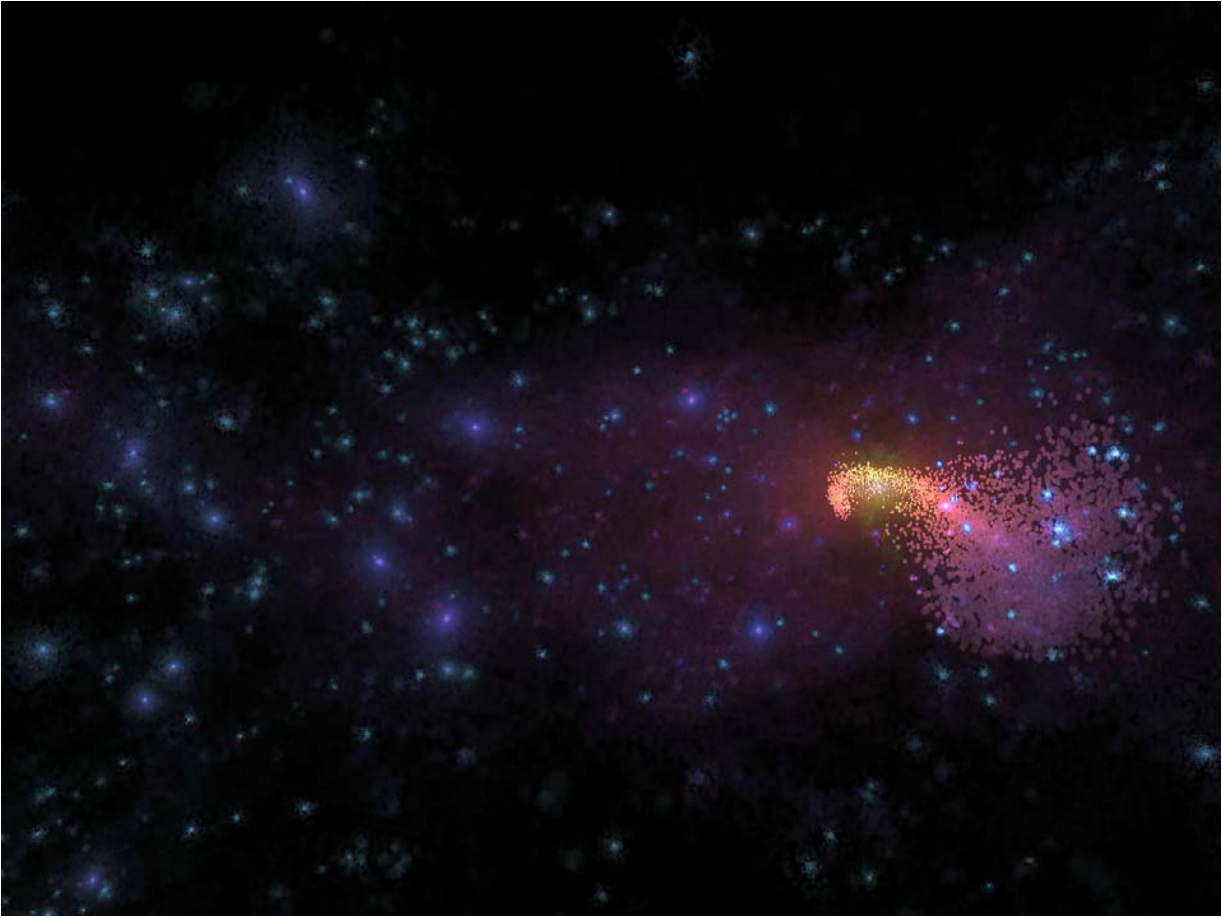

3. An example of highlighting objects of interest, in this case, during a collision.